\documentclass[epjCONF,onecolumn]{svjour}
\usepackage{graphics,amsmath}
\usepackage[varg]{txfonts} 
\usepackage[latin1]{inputenc}
\session-title{Conference Title, to be filled}
\begin{document}
\title{Similarity renormalization group evolution of $NN$ interactions \\ within a subtractive  renormalization scheme}
\author{V. S. Tim\'oteo\inst{1} \and S. Szpigel\inst{2} \and F. O. Dur\~aes\inst{2} }
\institute{Faculdade de Tecnologia, Universidade Estadual de Campinas\\ 13484-332 Limeira, SP, Brasil
\and Centro de Ci\^encias e Humanidades, Universidade Presbiteriana Mackenzie \\ 01302-907, S\~ao Paulo, SP, Brasil}
\abstract{
We apply the similarity renormalization group (SRG) approach to evolve a nucleon-nucleon ($NN$) interaction in leading-order (LO) chiral effective field theory (ChEFT), renormalized within the framework of the subtracted kernel method (SKM). We derive a fixed-point interaction and show the renormalization group (RG) invariance in the SKM approach. We also compare the evolution of $NN$ potentials with the subtraction scale through a SKM RG equation in the form of a non-relativistic Callan-Symanzik (NRCS) equation and the evolution with the similarity cutoff through the SRG transformation.
} 
\maketitle
\section{Introduction}
\label{intro}

The standard method for the non-perturbative renormalization of nucleon-nucleon ($NN$) interactions in the context of chiral effective field theory (ChEFT), which is inspired by Wilson's renormalization group \cite{wilson}, consists of two steps \cite{lepage,epelrev}. The first step is to solve the Lippmann-Schwinger (LS) equation with the $NN$ potential truncated at a given order in the ChEFT expansion, which consists of pion-exchange and contact interaction terms. This requires the use of a regularization scheme in order to overcome the ultraviolet divergences generated in the momentum integrals when such potentials are iterated. The most common approach used to regularize the LS equation is to introduce a sharp or smooth regularizing function that suppresses the contributions from the potential matrix elements for momenta larger than a cutoff scale, thus eliminating the divergences in the momentum integrals \cite{bira1}. The second step is to determine the renormalized strengths of the contact interactions, the so called low-energy constants (LEC's), by fitting a set of low-energy scattering data. Once the LEC´s are fixed, the LS can be solved to evaluate other observables.

Effective field theories and renormalization methods are based on the premise that physics at low-energy/long-distance scales is insensitive with respect to the details of the dynamics at high-energy/short-distance scales \cite{lepage}. In the case of ChEFT, the relevant high-energy effects for describing the low-energy observables can be captured in the scale-dependent LEC's of the $NN$ effective interactions. The $NN$ potential is considered correctly renormalized when the calculated observables are approximately independent of the cutoff in the range of validity of the ChEFT.  In the language of Wilson's renormalization group, this means that the LEC's must run with the cutoff in such a way that the scattering amplitude become (approximately) renormalization group invariant.

The renormalization group (RG) techniques have been successfully applied to analyze the scale dependence and the power counting scheme of $NN$ iteractions in the context of ChEFT \cite{birse1,birse2,birse3} and to derive phase-shift equivalent softer potentials from phenomenological $NN$ interactions by consistently integrating out high-momentum components, the so called $V_{low k}$ potentials \cite{vlow1,vlow2,vlow3}. Another RG approach that has been recently applied to evolve phenomenological and chiral effective field theory (ChEFT) $NN$ interactions to phase-shift equivalent softer forms is the similarity renormalization group (SRG) \cite{srg1,srg2,srg3}. The SRG formalism, developed by Glazek and Wilson \cite{wilgla1,wilgla2} and independently by Wegner \cite{wegner}, is a renormalization approach based on a series of continuous unitary transformations that evolve hamiltonians with a cutoff on energy differences. Viewing the hamiltonian as a matrix in a given basis, the similarity transformations suppress off-diagonal matrix elements as the cutoff is lowered, forcing the hamiltonian towards a band-diagonal form and effectively decoupling low-energy observables from high-energy degrees of freedom.

In this work we apply the SRG transformation to evolve an effective $NN$ interaction in leading-order (LO) ChEFT, derived within the framework of the subtracted kernel method (SKM) \cite{npa99,plb00,hepph01,plb05,npa07,ijmpe07}. The SKM is a renormalization scheme in which instead of using a cutoff regularizing function, the LS equation is regularized by performing subtractions in the kernel at a given energy scale, while keeping the original $NN$ interaction intact. A similar approach based on subtractive renormalization of the LS equation is described in Ref.~\cite{yang1,yang2}, although a momentum cutoff is also introduced to regularize the momentum integrals.

\section{Subtractive Renormalization}
\label{subren}

In the following, we show how to derive the subtracted kernel
LS equation \cite{npa99}. Here and in what follows we use units such that $\hbar=c=M=1$, where $M$ is the nucleon mass. 

Consider the formal LS equation for the $T$-matrix, written in operator form as
\begin{eqnarray}
T(E) &=& V + V~G_{0}^{+}(E)~T(E) = V \left[ 1+ G_{0}^{+}(E)~T(E) \right]\; ,
\label{LS}
\end{eqnarray}
\noindent
where $V$ is the interaction potential and $G_{0}^{+}(E)$ is the free Green's function for the two-nucleon system, given in terms of the free hamiltonian $H_0$ by
\begin{equation}
G_{0}^{+}(E) = [E - H_{0} + i \epsilon]^{-1} \; .
\end{equation}

Using Eq. (\ref{LS}), we can express the potential $V$ in terms of the $T$-matrix at a given energy scale $-\mu^2$,
\begin{eqnarray}
V &=& T(-\mu^2) \left[  1+G(-\mu^2)T(-\mu^2) \right]^{-1} \; .
\label{Vin}
\end{eqnarray}
\noindent
Putting Eq. (\ref{Vin}) back in Eq. (\ref{LS}), we obtain
\begin{eqnarray}
T(E) &=& T(-\mu^2)\left[1 + G_{0}^{+}(-\mu^2)~T(-\mu^2) \right]^{-1}  \nonumber \\
&+& T(E)G_{0}^{+}(E)T(-\mu^2) \left[1 + G_{0}^{+}(-\mu^2)T(-\mu^2) \right]^{-1}   \; .
\label{LS2}
\end{eqnarray}
\noindent
Now, we can multiply the entire equation by the term in the square brackets to find the subtracted kernel LS equation,
\begin{eqnarray}
T(E) &=& T(-\mu^2)
+ T(-\mu^2)
\underbrace{\left[ G_{0}^{+}(E) - G_{0}^{+}(-\mu^2) \right]}_{G^{(1)}(E;-\mu^2)}~T(E) \; ,
\label{SKLS}
\end{eqnarray}
\noindent
The Green's function after one subtraction, $G^{(1)}(E;-\mu^2)$, can be written as
\begin{eqnarray}
G^{(1)}(E;-\mu^2) &\equiv& \left[(-\mu^2-E)~ G_{0}^{+}(-\mu^2) \right]~G_{0}^{+}(E) = \frac{(-\mu^2-E)}{(-\mu^2-H_{0})}~G_{0}^{+}(E).
\label{G1}
\end{eqnarray}
\noindent
The subtracted kernel equation has the same structure as the formal LS equation Eq. (\ref{LS}) but with the potential $V$ replaced
by $T(-\mu^2)$ and the free Green's function, $G_{0}^{+}(E)$, replaced by $G^{(1)}(E;-\mu^2)$, which contains
one subtraction at the energy scale $-\mu^2$. For a pure contact interaction, the subtracted kernel LS equation produces a finite $T$-matrix due the presence of the form-factor in $G^{(1)}$. 

\subsection{Recursiveness and Multiple Subtractions}
\label{mult}

If we consider an interaction that contains stronger divergencies, like the ChEFT $NN$ interaction up to NNLO, we need more subtractions
to render the $T$-matrix finite. As the degree of the divergency increases, the kernel subtractions
can be performed recursively to obtain a finite $T$-matrix \cite{plb00}. The general LS equation for multiple recursive subtractions is given by
\begin{eqnarray}
T_n(E) &=& V^{(n)}(E;-\mu^2) + V^{(n)}(E;-\mu^2)G^{(n)}(E;-\mu^2)T(E) \; ,
\label{LSn}
\end{eqnarray}
\noindent
where the generic $n$-subtracted kernel is
\begin{eqnarray}
G^{(n)}(E;-\mu^2) &\equiv& \left[(-\mu^2-E)~ G_{0}^{+}(-\mu^2) \right]^{n}~G_{0}^{+}(E) = \frac{(-\mu^2-E)^n}{(-\mu^2-H_{0})^n}~G_{0}^{+}(E).
\label{Gn1}
\end{eqnarray}
\noindent
The recursive driving terms can be written as
\begin{eqnarray}
{V}^{(n)}(E;-\mu^2) &=&V^{(n-1)}(E;-\mu^2) + V^{(n-1)}(E;-\mu^2)~g^{(n)}(E;-\mu^2)~V^{(n)} \; ,
\label{dtVn}
\end{eqnarray}
\noindent
with
\begin{eqnarray}
g^{(n)}(E;-\mu^2) &\equiv& (-\mu^2-E)^{n-1}\left[G_{0}^{+}(-\mu^2) \right]^{n} = \frac{(-\mu^2-E)^{n-1}}{(-\mu^2-H_{0})^n} \; .
\label{Gn2}
\end{eqnarray}
\noindent
In order to obtain the $n$-subtracted $T$-matrix from Eq. (\ref{LSn}), first one needs to recursively
solve Eq. (\ref{dtVn}) for $V^{(n)}$ up to the number of subtractions required to regularize the considered interaction.

\subsection{Fixed-point Interactions and Renormalization Group Invariance}
\label{fprgi}

Given the driving term $V^{(n)}$, we can construct a fixed-point hamiltonian \cite{hepph01}, $H_{\cal R} = H_0 + V_{\cal R}$,
where $V_{\cal R}$ is the renormalized interaction. Both $H_{\cal R}$ and $V_{\cal R}$ are fixed-point operators, i.e. invariant with respect to the subtraction scale $-\mu^2$.

Replacing $V_{\cal R}$ in Eq. (\ref{LS}) we obtain the LS equation for the corresponding renormalized $T$-matrix,
\begin{eqnarray}
T_{\cal R}(E) = V_{\cal R} ~+~V_{\cal R}~G(E)~T_{\cal R}(E) \; .
\label{TRLS}
\end{eqnarray}
\noindent
By construction, the $T$-matrix obtained from the solution of this equation must be equivalent to that obtained from the solution of the $n$-subtracted kernel LS equation: $T_{\cal R}(E)=T_n(E)$. From this condition we obtain an integral equation which relates the renormalized interaction $V_{\cal R}$ to the driving term $V^{(n)}$,
\begin{eqnarray}
V_{\cal R} = {V}^{(n)}(E;-\mu^2) ~-~ {V}^{(n)}(E;-\mu^2)~~g_{\cal R}^{(n)}(E;-\mu^2)~~V_{\cal R} \; ,
\label{vrvn}
\end{eqnarray}
\noindent
with
\begin{eqnarray}
g^{(n)}_{\cal R} = \left[ 1 - \frac{(-\mu^2-E)^n}{(-\mu^2-H_{0})^n} \right] G_0^+(E) \; .
\end{eqnarray}

The subtraction scale $-\mu^2$ is arbitrary and so all observables should not depend on its choice. This condition is fulfilled by imposing the invariance of the $T$-matrix with respect to $-\mu^2$, which yields to a renormalization group equation in the form of a non-relativistic Callan-Symanzik (NRCS) equation,
\begin{eqnarray}
\frac{\partial}{\partial\mu^2} V^{(n)}(E;-\mu^2) &=& -V^{(n)}\frac{\partial G_{n}^{+}(E;-\mu^2)}{\partial\mu^2} V^{(n)}(E;-\mu^2) \; .
\label{CSEn}
\end{eqnarray}
\noindent
with the boundary condition $V^{(n)}(E;-\mu^2)|_{\mu \rightarrow {\bar \mu}}= V^{(n)}(E;-{\bar \mu}^2)$ imposed at some reference scale $\bar \mu$ where the physical information is supplied. This equation governs the evolution of the driving term $V^{(n)}(E;-\mu^2)$ with the subtraction scale $-\mu^2$ in such a way that the $T$-matrix remains invariant.

\begin{figure}[t]
\begin{center}
\resizebox{1.0\textwidth}{!}
{
  \includegraphics{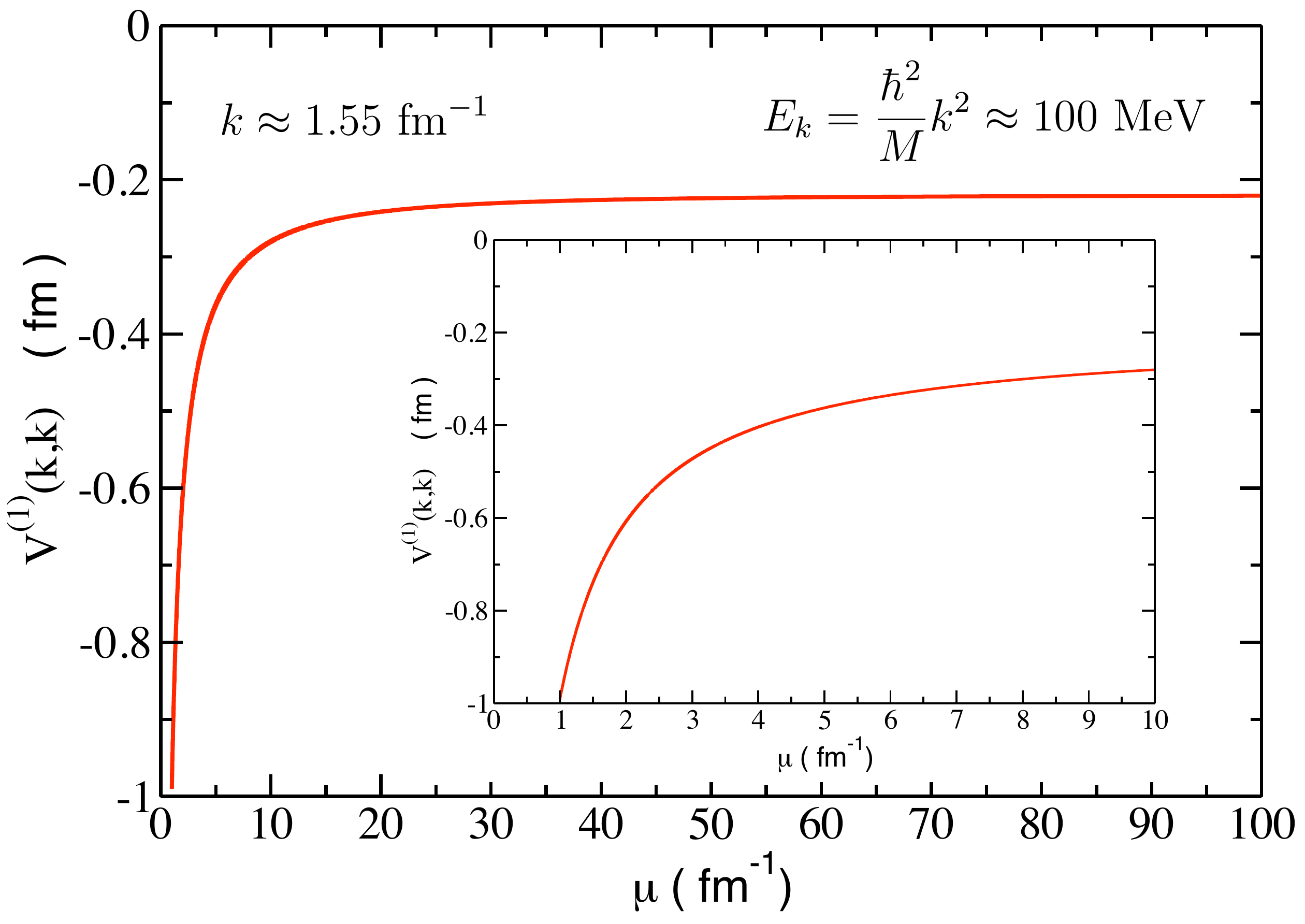}\hspace{0.5cm}
  \includegraphics{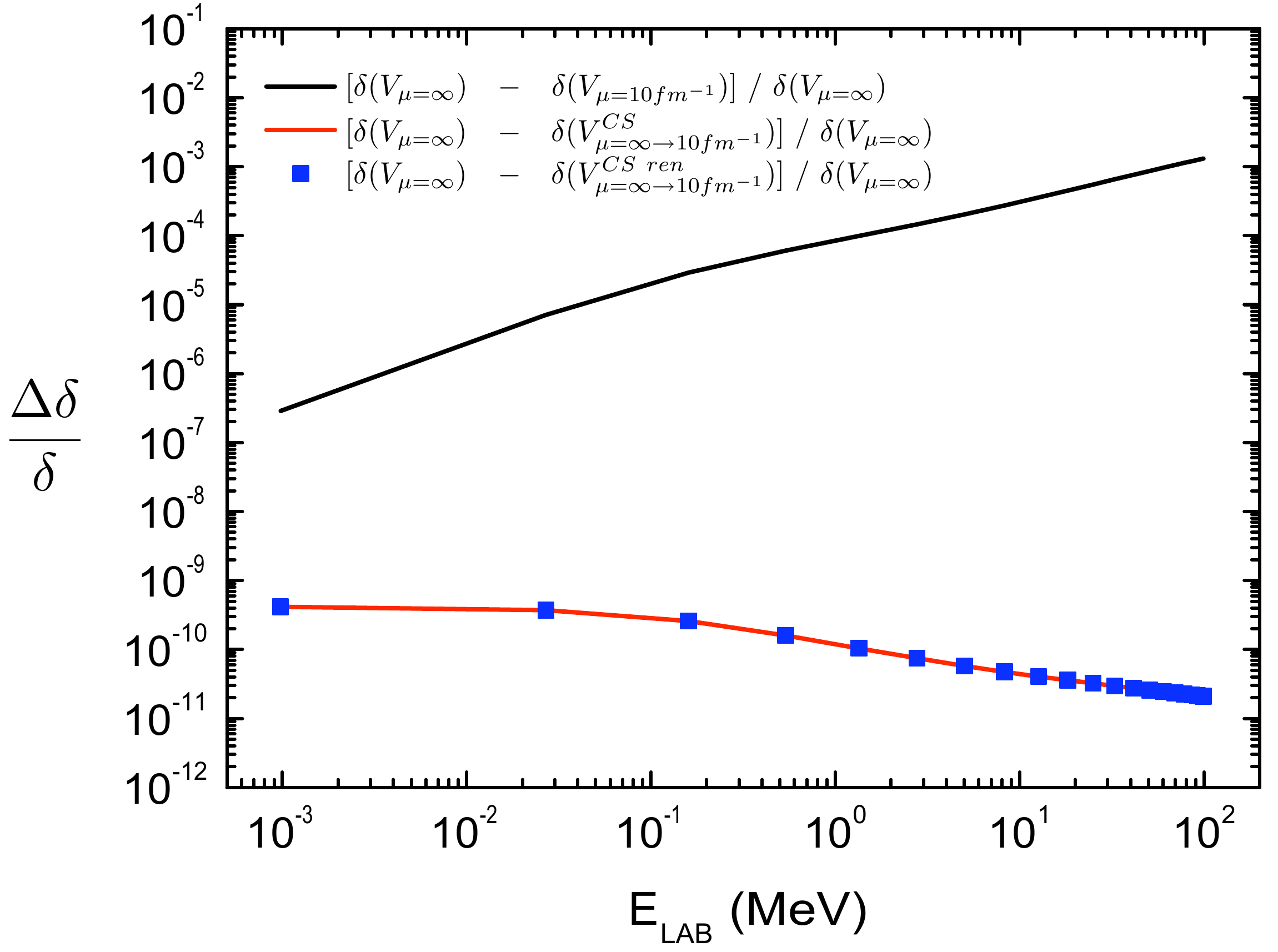}
}
\end{center}
\caption{Left panel: Evolution of $V^{(1)}(k,k)$ with the subtraction scale $-\mu^2$; Right panel: Relative differences between the phase-shifts in the $^1 S_0$ channel calculated at a scale $\mu=10~\rm{fm}^{-1}$ and at $\mu \rightarrow \infty$ in three distinct cases.}
\label{runrgi}
\end{figure}

In order to illustrate the renormalization group invariance in the SKM approach, we consider the LO ChEFT $NN$ interaction in the $^1S_0$ channel, which consists of the one-pion exchange potential (OPEP) plus a Dirac-delta contact interaction. The driving term for the corresponding one-subtracted LS equation ($n=1$) is given by
\begin{equation}
{V}^{(1)}(E;-\mu^2)= V_{\rm OPEP} + C_{0}(-\mu^2) \; ,
\label{driving1}
\end{equation}
\noindent
where $C_{0}(-\mu^2)$ is the renormalized strength of the contact interaction.

We solve Eq. (\ref{CSEn}) for the matrix elements $V^{(1)}(p,p')$ of the driving term projected in the $^1 S_0$ channel, using a partial-wave relative momentum space basis. The boundary condition is set at the reference scale ${\bar \mu}\rightarrow\infty$, where the renormalized strength of the contact interaction $C_{0}(-\mu^2)$ is fixed by fitting the experimental value of the scattering length in the $^1 S_0$ channel, $a_s=-23.7~{\rm fm}$. Once the strength $C_{0}(-\mu^2)$ is fixed, and so the matrix elements $V^{(1)}(p,p')$ of the driving term are known, we can compute the corresponding matrix elements of the renormalized interaction $V_{\cal R}(p,p')$ by solving Eq. (\ref{vrvn}) numerically. In the left panel of Fig. \ref{runrgi} we display the result obtained for the evolution of the diagonal matrix element $V^{(1)}(k,k)$ with the subtraction scale $-\mu^2$ for $k \equiv \sqrt{E_k}\simeq1.55~\rm{fm}^{-1}$. As one can observe, $V^{(1)}(k,k)$ is enhanced (becoming more attractive) as $\mu$ decreases. In the limit $\mu \rightarrow \infty$, the driving term becomes independent of $\mu$ and matches the corresponding renormalized potential.

Renormalization group invariance can be easily verified by evaluating the phase-shifts in the $^1 S_0$ channel as a function of the laboratory energy $E_{\rm LAB}$. In the right panel of Fig. (\ref{runrgi}) we show the relative differences between the phase-shifts calculated at $\mu=10~\rm{fm}^{-1}$ and at $\mu \rightarrow \infty$ in three distinct cases. The black line corresponds to the result obtained from the solution of the subtracted kernel LS equation with the driving term $V^{(1)}$ determined by simply fixing the strength of the contact interaction $C_{0}(-\mu^2)$ at $\mu=10~\rm{fm}^{-1}$ to fit the scattering length. The red line corresponds to the result obtained by evolving the driving term $V^{(1)}$ through the NRCS equation from $\mu \rightarrow \infty$ to $\mu=10~\rm{fm}^{-1}$. The blue squares correspond to the result obtained by solving the LS equation for $T_{\cal R}$ with the renormalized potential $V_{\cal R}$ determined from the driving term $V^{(1)}$ evolved through the NRCS equation. As one can observe, when the phase-shifts are evaluated with the unevolved driving term there is a residual dependence on the scale $\mu$ due to the fitting procedure used to fix the renormalized strength $C_0(-\mu^2)$. On the other hand, when the driving term is evolved through the NRCS equation the phase-shifts remain invariant, apart from relative numerical errors smaller than $10^{-9}$. One should note that the phase-shifts obtained from the LS equation for the renormalized $T$-matrix are also invariant, since the renormalized potential determined from the evolved driving term is a fixed-point operator.

\section{Similarity Renormalization Group for $NN$ Interactions}
\label{srg}

\subsection{Theoretical background}
\label{form}

The general formulation of the SRG approach was developed by Glazek and Wilson \cite{wilgla1,wilgla2} in the context of light-front hamiltonian field theory, aiming to obtain effective hamiltonians in which the couplings between high and low-energy states are eliminated, while avoiding artificial divergences due to small energy denominators.

Consider an initial hamiltonian in the center of mass frame for a system of two nucleons, which can be written in the form $H=T_{\rm rel}+V$, where $T_{\rm rel}$ is the relative kinetic energy and $V$ is the $NN$ potential. The similarity transformation is defined by a unitary operator designed to act on the hamiltonian and evolve it with a cutoff $\lambda$ on free energy differences at the interaction vertices,
\begin{eqnarray}
H_\lambda\equiv U(\lambda)\; H \; U^\dagger(\lambda) \equiv T_{\rm rel}+f_{\lambda}\;{\overline V}_{\lambda}\;,
\label{strans}
\end{eqnarray}
\noindent
where $f_{\lambda}$ is a similarity function and ${\overline V}_{\lambda}$ is called the reduced interaction. The unitarity condition satisfied by $U(\lambda)$ is given by:
\begin{eqnarray}
U(\lambda)\;  U^\dagger(\lambda)&\equiv&U^\dagger(\lambda)\; U(\lambda)  \equiv 1\;.
\label{unitarity}
\end{eqnarray}
\noindent
The similarity function $f_{\lambda}$ is a regularizing function which suppresses the matrix elements between states with free energy differences larger then the cutoff $\lambda$, such that the hamiltonian is driven towards a band-diagonal form as $\lambda$ is lowered. Usually, $f_{\lambda}$ is chosen to be a smooth function of the similarity cutoff $\lambda$. A simpler choice is to use a step function. The similarity transformation can be defined in terms of an anti-hermitian operator $\eta_{\lambda}$ which generates infinitesimal changes of the cutoff $\lambda$,
\begin{equation}
\eta_\lambda=U(\lambda)\; \frac{dU^\dagger(\lambda)}{d \lambda}=-\eta_{\lambda}^\dagger\;.
\label{etadef}
\end{equation}
\noindent
Using this definition and the unitarity of $U(\lambda)$ we can derive a first-order differential equation for the evolution of the hamiltonian,
\begin{eqnarray}
\frac{d H_{\lambda}}{d \lambda}= \left [ H_\lambda,\eta_\lambda \right]\; ,
\label{difstrans}
\end{eqnarray}
\noindent
with the boundary condition $H_\lambda |_{_{\lambda \rightarrow \infty}} \equiv H$.

In the application of the SRG described in this work, we employ the formulation developed by Wegner \cite{wegner}, based on a flow equation that governs the unitary evolution of the hamiltonian
\begin{equation}
\frac{d H_s}{ds}=[\eta_s,H_s]
\label{wegner1}\; ,
\end{equation}
\noindent
with the boundary condition $H_s |_{_{s \rightarrow 0}} \equiv H$. The flow parameter $s$ has dimensions of $({\rm energy})^{-2}$ and ranges from $0$ to $\infty$. In terms of a similarity cutoff $\lambda$, here with dimensions of momentum, the flow parameter is given by the relation $s=\lambda^{-4}$.

Wegner's flow equation is analogous to Eq.(\ref{difstrans}), but the specific form $\eta_s=[G_s,H_s]$ is chosen for the anti-hermitian operator that generates the unitary transformation, which gives
\begin{equation}
\frac{d H_s}{ds}=[[G_s,H_s],H_s]\; .
\label{wegner2}
\end{equation}
\noindent
Such a choice for the generator $\eta_s$ corresponds in the Glazek-Wilson formulation to the choice of a gaussian similarity function $f_{\lambda}$ with uniform width $\lambda$ . The operator $G_s$ defines the generator $\eta_s$ and so specifies the flow of the hamiltonian. Wegner's choice in the original formulation is the full diagonal part of the hamiltonian in a given basis, $G_s={\rm diag}(H_s)$. A simpler choice is to use the free hamiltonian, $G_s=T_{\rm rel}$, which gives the generator $\eta_s=[T_{\rm rel},H_s]$

\subsection{SRG Evolution in the SKM Approach}
\label{srglo}

Using the generator $\eta_s=[T_{\rm rel},H_s]$, Wegner's flow equation for the SRG evolution of the $NN$ potential matrix elements is given by
\begin{eqnarray}
\frac{dV_{s}(p,p')}{ds}=-(p^2-p'^2)^2 \; V_{s}(p,p')+\frac{2}{\pi} \int_{0}^{\infty}dq \; q^2~(p^2+p'^2-2 q^2)~ V_{s}(p,q)\; V_{s}(q,p').
\label{flowNN}
\end{eqnarray}
\noindent
For simplicity, we are using $V_{s}(p,p')$ as a short notation for the projected $NN$ potential matrix elements $V^{~(JLL'S;I)}_{s}(p,p')$ in a partial-wave relative momentum space basis, $|\;q{(LS)J;I} \;\rangle$, with normalization
\begin{equation}
1=\frac{2}{\pi} \int_{0}^{\infty}dq \; q^2 \; |\;q{(LS)J;I} \;\rangle \;\langle \; q{(LS)J;I}\;|,
\label{PWnorm}
\end{equation}
\noindent
where the indexes $J$, $L(L')$, $S$ and $I$ respectively denote the total angular momentum, the orbital angular momentum, the spin and the isospin quantum numbers of the $NN$ state. For non-coupled channels ($L=L'=J$), such as the singlet $^1 S_0$, the $NN$ potential matrix elements $V_{s}(p,p')$ are simply given by $V_{s}(p,p')=V^{(JJJS;I)}_{s}(p,p')$. For coupled channels ($L,L'=J \pm 1$), such as the triplet $^3 S_1 - ^3 D_1$, the $V_{s}(p,p')$ represent $2 \times 2$ matrices of matrix elements for the different combinations of $L$ and $L'$, 
\begin{eqnarray}
 V_{s}(p,p')
  \equiv \begin{pmatrix}
    V^{(JLLS;I)}_{s}(p,p')  & V^{(JLL'S;I)}_{s}(p,p') \\
    V^{(JL'LS;I)}_{s}(p,p')  & V^{(JL'L'S;I)}_{s}(p,p')
  \end{pmatrix} \;.
\end{eqnarray}
\noindent
As a consequence of the choice of the SRG transformation generator $\eta_s=[T_{\rm rel},H_s]$ each interaction channel evolves with the cutoff $\lambda=s^{-1/4}$ independently of the other channels \cite{UCOM}.

\begin{figure}[t]
\begin{center}
\resizebox{0.90\textwidth}{!}
{
  \includegraphics{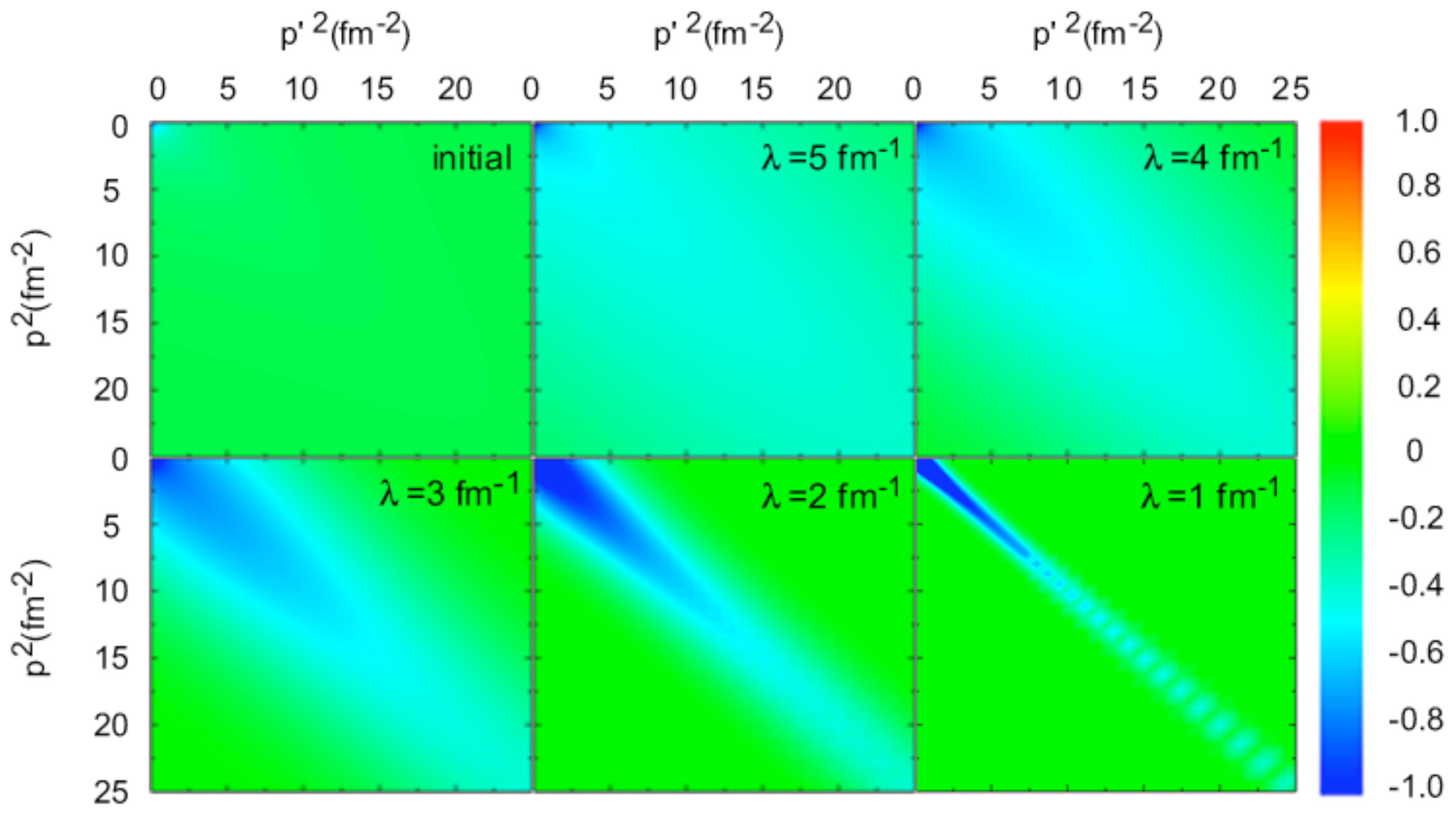}
}
\end{center}
\caption{(Color online) SRG evolution of the SKM-LO ChEFT potential in the $^1 S_0$ channel (in units of ${\rm fm}$).}
\label{losrg}
\end{figure}

We solve Eq. (\ref{flowNN}) numerically, obtaining an exact (non-perturbative) solution for the evolution of the SKM-LO ChEFT potential in the $^1 S_0$ channel. The relative momentum space is discretized on a grid of $N$ gaussian integration points (we have used 200 mesh points), leading to a system of $N^2$ non-linear first-order coupled differential equations which is solved using an adaptative fifth-order Runge-Kutta algorithm. The boundary condition is set at $s=0$ ($\lambda \rightarrow \infty$), such that the initial potential is given by the fixed-point renormalized interaction $V_{\cal R}(p,p')$ derived through the SKM scheme. As one can observe from Fig. (\ref{losrg}), the off-diagonal matrix elements are systematically suppressed as the similarity cutoff $\lambda$ is lowered, such that the potential is driven towards a band-diagonal form.

\begin{figure}[t]
\begin{center}
\resizebox{0.90\textwidth}{!}
{
  \includegraphics{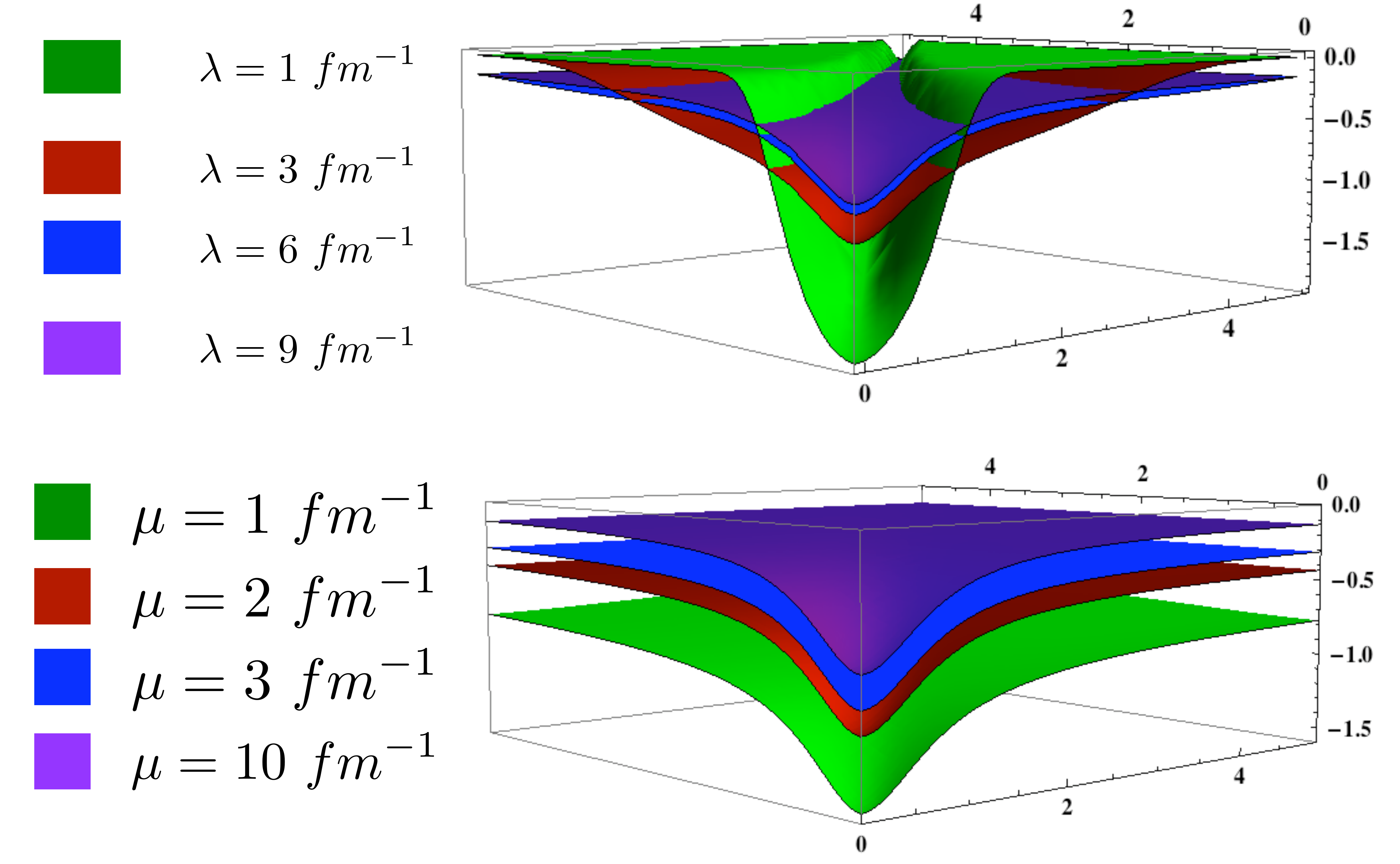}
}
\end{center}
\caption{Top panel: SRG evolution for the renormalized potential; Bottom panel: Evolution of the driving term through the NRCS equation.
Units are $fm$ for the potential and $fm^{-1}$ for the momenta.}
\label{srgcs}
\end{figure}

In Fig. \ref{srgcs} we compare the SRG evolution of the matrix elements $V_{\cal R}(p,p')$ for the fixed-point renormalized interaction with the similarity cutoff $\lambda$ (top) and the evolution through the NRCS equation of the matrix elements $V^{(1)}(p,p')$ for the driving term with the subtraction scale $\mu$ (bottom). As one can observe, the NRCS evolution preserves the global shape of the driving term. On the other hand, the SRG evolution completely modifies the shape of the renormalized interaction.

\begin{figure}[h]
\begin{center}
\resizebox{0.90\textwidth}{!}
{
  \includegraphics{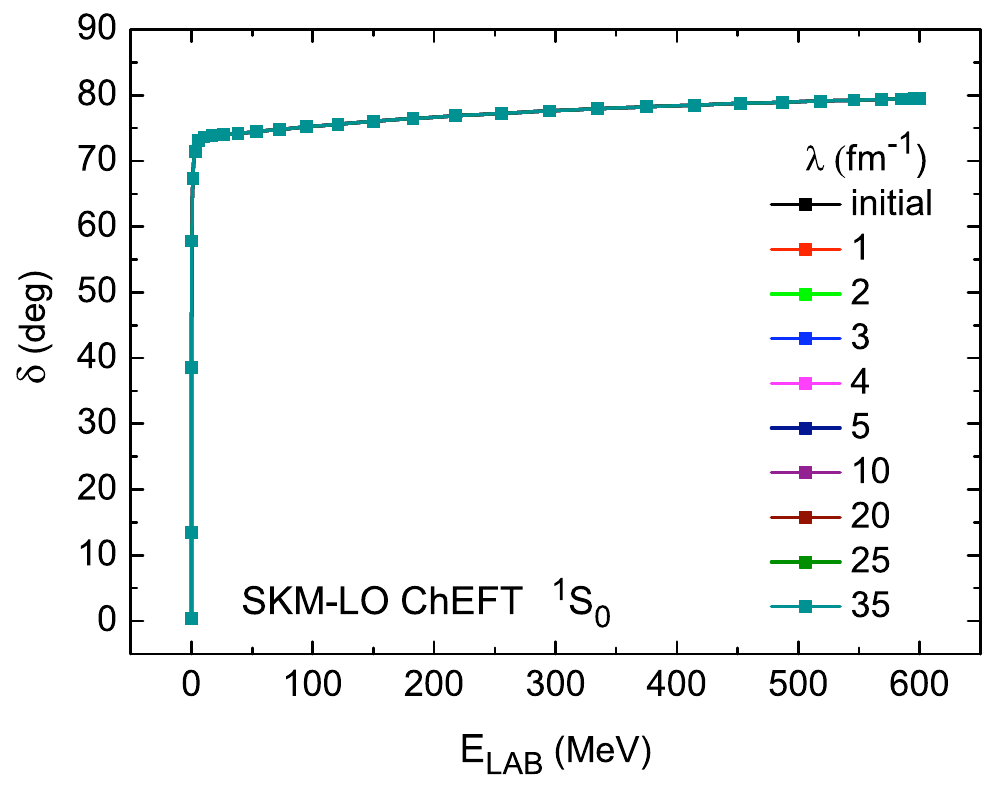} \hspace{0.8cm} \includegraphics{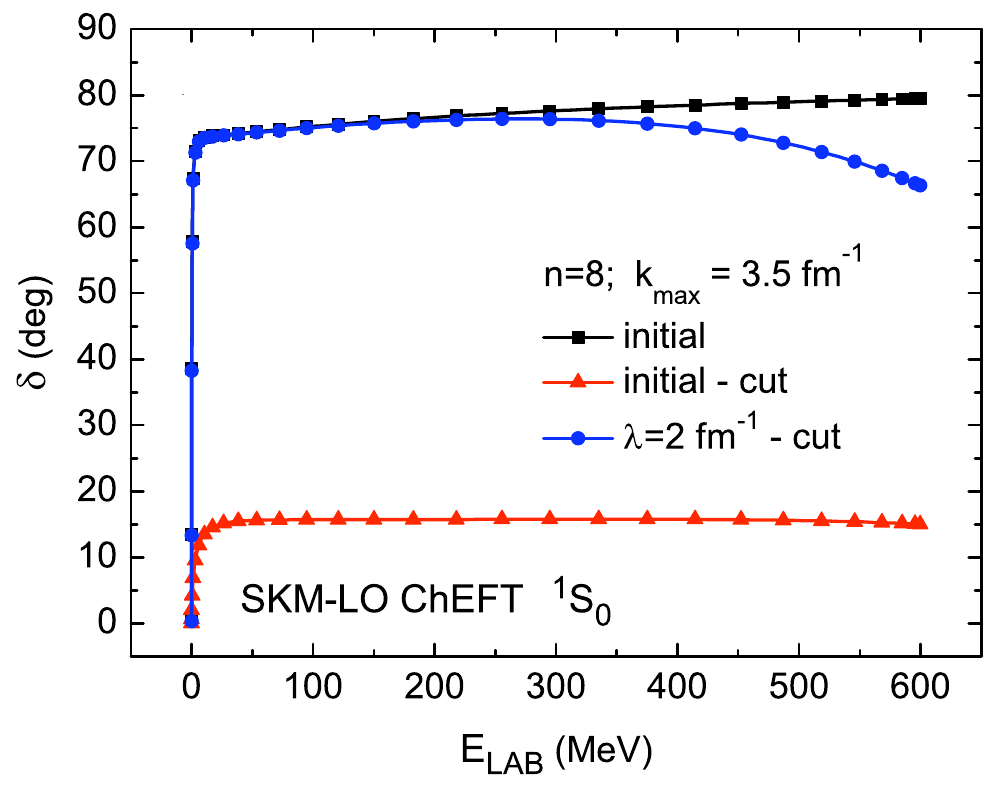}
}
\end{center}
\caption{Left panel: Phase-shifts in the $^1 S_0$ channel as a function of the laboratory energy; Right panel: Test for the decoupling between low- and high-momentum components.}
\label{psdec}
\end{figure}

In the left panel of Fig. (\ref{psdec}) we show the phase-shifts in the $^1 S_0$ channel obtained for the initial fixed-point renormalized interaction $V_{\cal R}(p,p')$ and for the corresponding SRG potentials evolved up to several values of the similarity cutoff $\lambda$. As expected for a unitary transformation, the results are the same (apart from relative numerical errors smaller than $10^{-9}$).

We now test for the decoupling of low-energy observables from high-energy degrees of freedom following the analysis introduced by Bogner et al. \cite{srg2,srg3}, which consists in applying an exponential function to the SRG potential that suppresses contributions from matrix elements $V_{s}(p,p')$ with $p,p'$ larger than a given momentum $k_{\rm max}$,
\begin{equation}
V^{(k_{\rm max}, n)}_{s}(p,p')={\rm exp}[-(p^2/k_{\rm max}^2)^n]\; V_{s}(p,p') \;{\rm exp}[-(p'^2/k_{\rm max}^2)^n] \;,
\label{smooth}
\end{equation}
\noindent
In the right panel of Fig. (\ref{psdec}) we show the phase-shifts obtained by cutting the initial potential $V_{\cal R}(p,p')$ and the SRG potential evolved to a similarity cutoff $\lambda=2 \; {\rm fm}^{-1}$ at $k_{\rm max}=3.5 \; {\rm fm}^{-1}$, with $n=8$. As one can observe, the phase-shifts obtained for the cut initial potential $V_{\cal R}(p,p')$ are completely modified in comparison to those for the uncut potential. For the cut SRG potential, the phase-shifts remain practically unchanged at low energies. Therefore, an explicit decoupling between the low- and high-momentum components is verified for the SRG evolved potential.

\section{Summary and Concluding Remarks}
\label{conc}

We have investigated the similarity renormalization group evolution (SRG) of $NN$ interactions in the framework of the subtracted kernel method (SKM), a renormalization scheme based on subtractions performed in kernel of the scattering equation. 

Considering a simple example, the LO ChEFT $NN$ interaction in the $^1S_0$ channel, we have shown that a fixed-point renormalized interaction and renormalization group invariant phase-shifts can be obtained from the subtracted kernel scattering equation, provided the driving term is evolved with the subtraction scale through a renormalization group equation in the form of a non-relativistic Callan-Symanzik (NRCS) equation. We have solved Wegner's flow equation numerically to obtain a non-perturbative solution for the SRG evolution of the fixed-point renormalized interaction. By calculating the phase-shifts, we have verified the unitarity of the similarity transformation. By cutting the SRG potential at a given momentum using an exponential function, we have verified the decoupling of low-energy observables from the hig-energy degrees of freedom. 

The next step, to be implemented in a future work, is to consider the the SRG evolution of $NN$ interactions up to higher-orders in ChEFT and other partial-wave channels.

\section*{Acknowledgements}
\begin{acknowledgement}
This work was supported by CNPq, FAPESP and Instituto Presbiteriano Mackenzie through Fundo Mackenzie de Pesquisa. V. S. T. would like to thank Prof. Evgeny Epelbaum for the hospitality at the IKP/FZJ during the week right before FB19.
\end{acknowledgement}

\end{document}